\documentclass[twocolumn,conference,10pt]{IEEEtran}

\usepackage{amsmath, amssymb,graphicx,cite}
\usepackage{srcltx}
\usepackage{epsfig,amsfonts}
\usepackage{graphicx,cite,amssymb,amsmath}
\usepackage{color}
\usepackage[nolist]{acronym}
\usepackage{psfrag}
\usepackage{perso}
\ifCLASSOPTIONcompsoc
\usepackage[caption=false,font=normalsize,labelfont=sf,textfont=sf]{subfig}
\else
\usepackage[caption=false,font=footnotesize]{subfig}
\fi
\usepackage{framed}
\IEEEoverridecommandlockouts

\begin{document}
\begin{acronym}
\acro{LT}{Luby Transform}
\acro{BP}{belief propagation}
\acro{LRFC}{linear random fountain code}
\acro{ML}{maximum likelihood}
\acro{GE}{Gaussian elimination}
\acro{RSD}{robust soliton distribution}
\acro{SA}{simulated annealing}
\end{acronym}
\newcommand{\figw}{1\columnwidth}

\title{LT Code Design for Inactivation Decoding}

\author{Francisco L\'azaro Blasco$^*$, Gianluigi Liva$^*$, Gerhard Bauch$^\dagger$\\
$^*$ Institute of Communications and Navigation\\
DLR (German Aerospace Center), Wessling, Germany 82234\\
$^\dagger$ Institute for Telecommunication\\
Hamburg University of Technology, Hamburg, Germany\\
\normalsize Email: {\tt Francisco.LazaroBlasco@dlr.de, Gianluigi.Liva@dlr.de, Bauch@tuhh.de}
\thanks{This work will be presented at the IEEE Information Theory Workshop (ITW) 2014, Hobart, Australia}

\thanks{\copyright 2014 IEEE. Personal use of this material is permitted. Permission
from IEEE must be obtained for all other uses, in any current or future media, including
reprinting /republishing this material for advertising or promotional purposes, creating new
collective works, for resale or redistribution to servers or lists, or reuse of any copyrighted
component of this work in other works}
}

\maketitle

\thispagestyle{empty}


\begin{abstract}
We present a simple model of inactivation decoding for \acs{LT} codes which can be used to estimate the decoding complexity as a function of the LT code degree distribution. The model is shown to be accurate in variety of settings of practical importance. The proposed method allows to perform a numerical optimization on the degree distribution of a \acs{LT} code aiming at minimizing the number of inactivations
required for decoding.
\end{abstract}

\thispagestyle{empty}

\setcounter{page}{1}


{\pagestyle{plain} \pagenumbering{arabic}}


\section{Introduction}\label{sec:Intro}

Fountain codes \cite{byers02:fountain} are a class of erasure correcting codes which can generate an endless number of encoded symbols. This feature makes them very useful when the erasure rate of the communication channel is not known. Fountain codes are also a very efficient solution for reliable multicast/broadcast transmissions in which a transmitter wants to deliver an object (file) to a set of receivers. Reliable multicasting is of special interest in wireless systems due to the broadcast nature of the transmission medium. For example, in our case we are interested in delivering a file via satellite to a set of ships on high seas.

The first class of practical fountain codes, \ac{LT} codes,  were introduced in \cite{luby02:LT} together with an efficient (iterative) \ac{BP} erasure decoding algorithm exploiting a sparse graph representation of the codes.  Raptor codes \cite{shokrollahi06:raptor} were introduced as an extension of \ac{LT} codes which consists of a serial concatenation which uses a \ac{LT} code as an inner code and an outer code which is normally chosen to be a high rate erasure correcting code.
\ac{BP} decoding of \ac{LT} codes is very efficient for long block lengths. However, the performance of \ac{BP} degrades for moderate and short block lengths.  In \cite{shokrollahi2005systems} inactivation decoding for \ac{LT} codes was introduced as an efficient ML decoding algorithm having manageable complexity for moderate/small block lengths. Inactivation decoding is widely used in practice (an exemplary case is the standard in \cite{MBMS12:raptor}). However, most of the analyses of \ac{LT} and Raptor codes focus on \ac{BP} decoding (see e.g. \cite{pakzad2006design}, \cite{maneva2006lt}). An exception is the work in \cite{mahdaviani2012raptor}, where the authors derived analytically some degree distributions optimized for inactivation decoding. In this work we study inactivation decoding for \ac{LT} codes. First we present a novel method which is able to accurately estimate the expected number of inactivations required by inactivation decoding for a given \ac{LT} code. This method is then embedded into a numerical optimization algorithm which searches for output degree distributions which minimize the number of inactivations. In contrast to the work in \cite{mahdaviani2012raptor} our algorithm allows  to freely set the average output degree of the distribution as well as to introduce arbitrary constraints on the code design.

The paper is organized as follows. In Section  \ref{sec:inact} we present how inactivation decoding works. In Section \ref{sec:model} we introduce the method to predict the complexity of inactivation decoding of \ac{LT} codes. Section \ref{sec:dist} describes the numerical optimization algorithm and provides examples of \ac{LT} code design. Finally we present the conclusions to our work in Section \ref{sec:Conclusions}.

\section{Inactivation Decoding of LT codes}\label{sec:inact}

\begin{figure}
        \centering
        \subfloat[Structure of $\tilde G$ after the triangularization procedure.]
        {\includegraphics[width=0.45\columnwidth,draft=false]{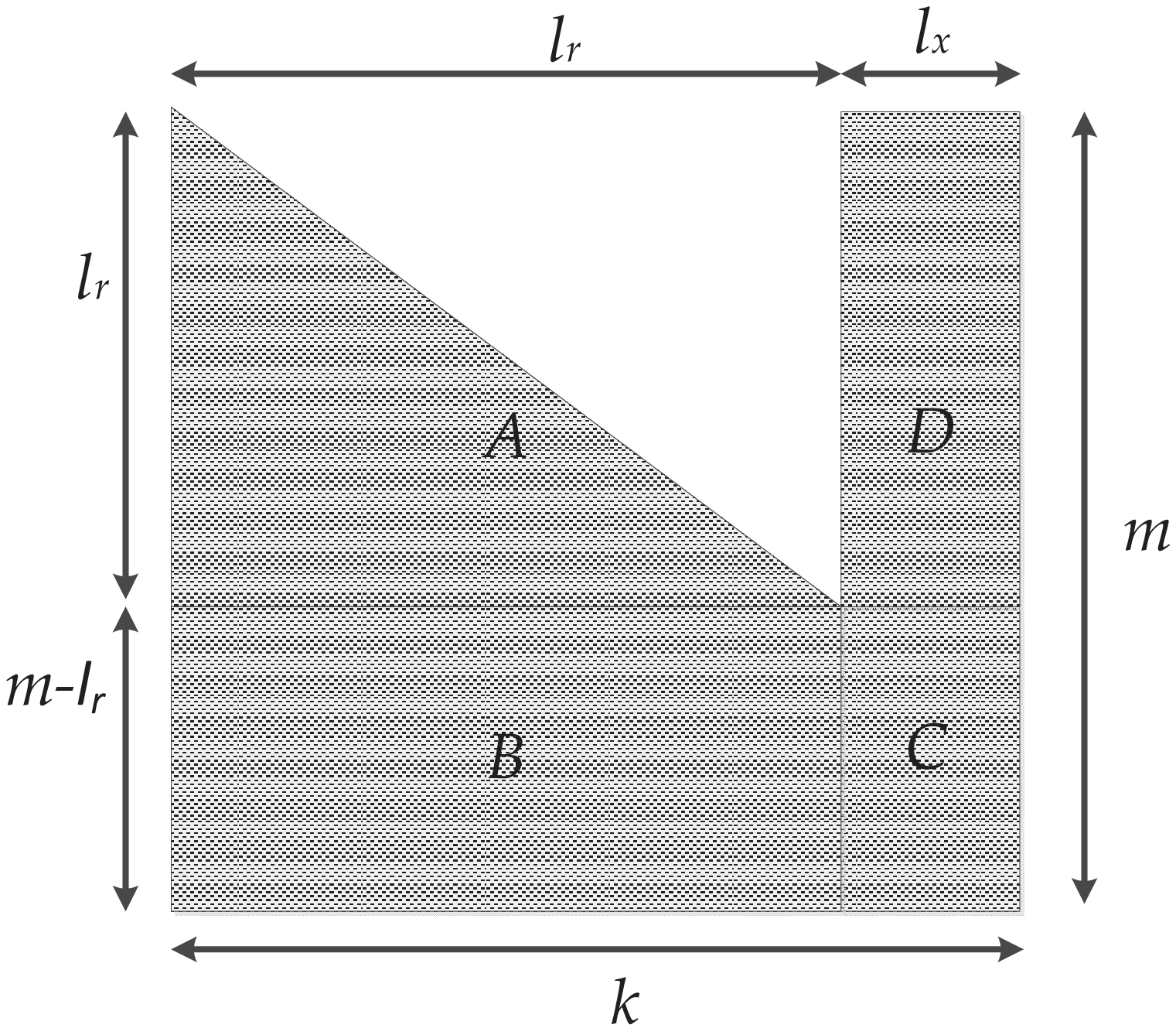}}
        \label{fig:piv_a}
        \qquad
        \subfloat[Structure of $\tilde G$ after the zero matrix procedure.]
        {\includegraphics[width=0.45\columnwidth,draft=false]{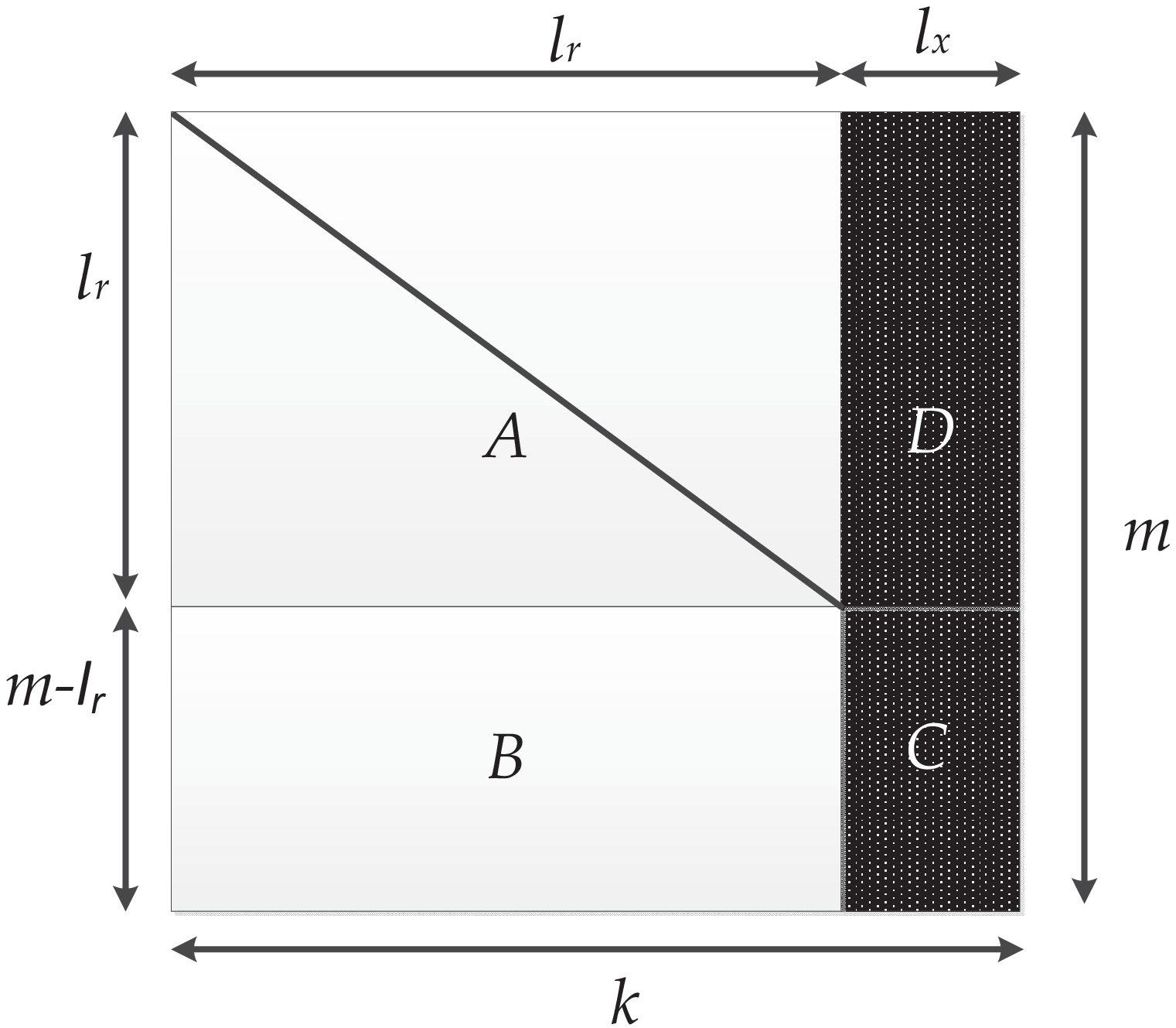}}
        \label{fig:piv_b}
        \caption{Triangulization and zero matrix procedure steps of inactivation decoding.}
\end{figure}

We consider a binary \ac{LT} code with $k$ input symbols $\mathbf{u}=(u_1, u_2, \ldots, u_k)$. The output degree distribution which defines the \ac{LT} code will be denoted as $\Omega= \{ \Omega_1, \Omega_2,\Omega_3, \hdots \Omega_{d_{\mathrm{max}}}\}$ where for the maximum degree we have $d_{\mathrm{max}} \leq k$. Assume the receiver has collected $m$ output symbols $\mathbf{c}=(c_1, c_2, \ldots, c_m)$. The relative receiver overhead is denoted by $\epsilon = 1 - m/k$. The decoder will have to solve the system of equations
\begin{equation}
\mathbf{c} =\mathbf{u} \mathbf{G}^T
\end{equation}
with $\mathbf{G}$ being the  $m \times k$ binary matrix defining the relation between the input and the output symbols.
For \ac{LT} codes, the matrix $\mathbf{G}$ is sparse. Efficient \ac{ML} decoding can be performed by exploiting the sparse nature of $\mathbf{G}$ through the following steps:
\begin{enumerate}
  \item {\emph{Triangularization.} $\mathbf{G}$ is put in an approximate lower triangular form. At the end of this process we are left with lower triangular matrix $\mathbf{A}$ and matrices $\mathbf{B}$,
  $\mathbf{C}$, and $\mathbf{D}$ which are sparse as shown in Fig.~\ref{fig:piv_a}. This process consists of column and row permutations.}
  \item {\emph{Zero matrix procedure.} The matrix $\mathbf{A}$ is put in a diagonal form and matrix $\mathbf{B}$ is zeroed out through row sums. As a consequence matrices $\mathbf{C}$ and $\mathbf{D}$ may become dense. The structure of $\mathbf{G}$ at the end of this procedure is shown in Fig.~\ref{fig:piv_b}.}
  \item {\emph{\ac{GE}}. \ac{GE} is applied to solve the systems of equations $\tilde{\mathbf{c}} =\tilde{\mathbf{u}} \mathbf{C}^T$, where $\tilde{\mathbf{u}}=(\tilde u_1, \tilde u_2,..., \tilde u_{l_x})$ are called \emph{reference variables} (associated with the rightmost columns of the matrix in Fig.~\ref{fig:piv_b}) and  $\tilde{\mathbf{c}}= (\tilde c_1, \tilde c_2,\ldots, \tilde c_{m-l_r})$ are $m-l_r$ known terms associated with the last $m-l_r$ of the matrix in Fig.~\ref{fig:piv_b} which depend only on the reference variables.}
  \item {\emph{Back-substitution.} Once the values of the reference variables $\tilde u_1, \tilde u_2,..., \tilde u_{l_x}$ has been determined, back-substitution is applied to compute the values of the remaining variables in $\mathbf{u}$.}
\end{enumerate}

\begin{figure}[h]
\begin{center}
\includegraphics[width=0.6\columnwidth,draft=false]{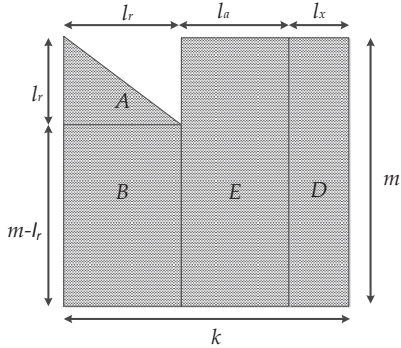}
\centering \caption{Structure of $\mathbf{G}$ at decoding step $q= l_r +l_x$. }
\label{fig:piv_c}
\end{center}
\end{figure}

Note that decoding is successful only if the rank of the sub-matrix $\mathbf C$ equals the number of reference variables, $l_x$.
In order to characterize inactivation decoding it is useful to move to a bipartite graph representation. In this representation output symbols will be denoted by squares nodes and input symbols by circles. As a consequence, each output symbol node will correspond to a row of the matrix $\mathbf G$ and each input symbol node will correspond to a column of the matrix $\mathbf G$. An output symbol node of degree $d$ will be connected with an edge to the $d$ input symbol nodes whose linear combination generates the output symbol.  At the beginning of the decoding all input and output symbol nodes are marked as \emph{active}. During the triangularization procedure, at each step the decoder marks an active input symbol node as either \emph{resolvable} or \emph{inactive}. An output symbol node is active as long as it has one or more active neighbours. The resolvable input symbol nodes correspond to the columns of matrix $\mathbf A$, whereas the inactive input symbol nodes correspond to the columns of $\mathbf D$. We further define the active degree of an input or output node as the number of active neighbors of the node.
Let us assume the decoder is at step $j=l_r+l_x$, being  $l_r$ the number of input symbol nodes marked as resolvable and $l_x$ the number of input symbol nodes marked as inactive (see Fig.~\ref{fig:piv_c}). We have that $l_a=k-l_r-l_x$ is the number of active input symbol nodes. At this stage the decoder operates as follows:
\begin{itemize}
  \item {The decoder tries to find an output symbol node (row) $\tilde c$  with only one active neighbor.
      \begin{itemize}
        \item {If such an output symbol exists, this symbol its only neighbor $u_x$ are marked as resolvable. This decreases the active degree of the output symbols which have $u_x$ as neighbor.}
        \item {If such an output symbol does not exist, an inactivation takes place, i.e. the decoder marks one of the $l_a$ active input symbol nodes as inactive.}
      \end{itemize}
      }
  \item {The decoder moves to step $j+1$.}
\end{itemize}
After $k$ steps all input symbols are either inactive or resolvable. After the zero out procedure, \ac{GE} is used to solve the systems of equations $\tilde{\mathbf{c}} =\tilde{\mathbf{u}} \mathbf{C}^T$. This step drives the complexity of decoding since \ac{GE} on a $n \times n$ matrix requires $\mathcal O(n^3)$ operations. Therefore, the complexity of inactivation decoding is dominated by the number of reference variables, $l_x$.
In the following, a way to compute the average number of inactivations needed at the decoder will be derived, which will depend on the degree distribution $\Omega$.

For the inactivation step,  different strategies can be applied to select the symbol to be inactivated (see e.g. \cite{paolini2012}, \cite{shokrollahi2005systems}). We consider two different inactivation techniques. The first strategy, \emph{random inactivation} consists simply of selecting uniformly at random the input symbol node to be inactivated. In the second strategy, \emph{maximum active degree inactivation},  the input symbol with maximum active degree is inactivated.

\section{A Model for Random Inactivation Decoding}\label{sec:model}
\begin{figure}
\begin{center}
\includegraphics[width=0.85\figw]{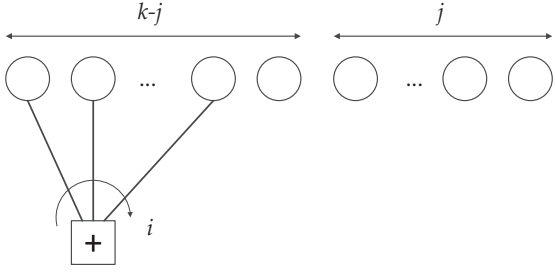}
\centering \caption{Output symbol belonging to $\mathcal{R}_i^{(j)}$. At step $j$, $k-j$ input symbols are still active. The symbol has $i$ edges going to active input symbols.}
\label{fig:graph}
\end{center}
\end{figure}

In this section we present a model to predict the average number of inactivations needed to decode as a function of the degree distribution $\Omega$, the input block size $k$ and the overhead $\epsilon$ under random inactivation.  We will denote as $i$-th output ripple at step $j$ of the algorithm, $\mathcal{R}_i^{(j)}$, the set of output symbol nodes of active degree $i$ when $k-j$ input symbols are still active (see Fig.~\ref{fig:graph}).  $R_i^{(j)}$ shall denote the cardinality of $\mathcal{R}_i^{(j)}$.We shall assume that an output symbol chooses its neighbors without replacement, in other words, we do not allow output symbols to throw more than one edge to the same input symbol.

The algorithm is based on the assumption that $R_i^{(j)}$ follows a binomial distribution with parameters $m^{(j)}$ and $p_i^{(j)}$, $\mathcal B (m^{(j)}, p_i^{(j)})$, where $m^{(j)}$ represents the number of active output symbols at step $j$ and $p_i^{(j)}$ represents the probability that one of the output symbols at step $j$ belongs to the $i$-th ripple. The assumption showed to be very accurate through extensive Monte Carlo simulations. According to the assumption,  we  have that $R_i^{(0)}$ initially follows a binomial distribution $\mathcal B (m, \Omega_i)$, i.e.
\begin{equation*}
\Pr (R_i^0 = q) = \binom {m}{q} {\Omega_i}^q (1-\Omega_i)^{m-q}.
\end{equation*}

Let us consider an output symbol $c_l$ which belongs to $\mathcal{R}_i^{(j)}$, $i>1$.  In other words, at step $j$ $c_l$ has $i$ neighbors among the $k-j$ unresolved symbols (see Fig.~\ref{fig:graph}).  The probability that $c_l$ leaves the $i-th$ ripple at step $j+1$, $\chi_i^{j+1}$, is the probability that one of these $i$ neighbors stops being active and becomes either resolvable or inactive. Under the no replacement assumption this probability takes the value
\begin{equation*}
\chi_i^{j+1} = \frac{i}{k-j}.
\end{equation*}
Recalling our assumption  that $R_i^{(j)} \sim \mathcal B \left(m^{(j)}, p_i^{(j)}\right)$,
the expected number of symbols leaving the $i$-th  ripple, $i>1$, at step $j+1$ will be
\begin{align*}
N_{i}^{j+1} &= \mathsf{E} \left[R_{i}^{(j)} \chi_{i}^{j+1}\right] \nonumber \\ &= \chi_{i}^{j+1} \mathsf{E} \left[R_{i}^{(j)}\right] \nonumber \\
&= \chi_{i}^{j+1} m^{(j)} p_i^{(j)}.
\end{align*}
For the case $i=1$ the number of symbols leaving the first ripple will be
\begin{align*}
N_1^{(j+1)} &= \mathsf{E} \left[ 1+ (R_{1}^j-1) \frac{1}{k-j} \right]
\end{align*}
when no inactivation takes place, and
\begin{align*}
N_1^{(j+1)} &= 0
\end{align*}
when an inactivation is performed.
Since an inactivation occurs when $R_{1}^j = 0$, we have that
\begin{align*}
N_1^{(j+1)} &=  \left( 1-\frac{1}{k-j} \right) \Pr (R_{1}^j = 0) +\frac{1}{k-j} \mathsf{E} \left[ R_{1}^j \right]\\
 &= \left( 1-\frac{1}{k-j} \right) \left(1-p_1^{(j)}\right)^{m^{(j)}} + \frac{1}{k-j} m^{(j)} p_1^{(j)}.
\end{align*}
Analogously, the expected number of symbols entering the $i$-th ripple at step $j+1$ corresponds to the number symbols which leave the $i+1$-th ripple,
\begin{equation*}
N_i^{j+1} = \mathsf{E} \left[ R_{i+1}^{(j)} \chi_{i+1}^{j+1} \right] = \chi_{i+1}^{j+1} m^{(j)} p_{i+1}^{(j)}.
\end{equation*}
The expected number of active output symbols in the graph at step $j+1$ can be computed recursively as
\begin{equation*}
m^{(j+1)} = m^{(j)} - {N}_1^{(j+1)},
\end{equation*}
and $p_{i+1}^{(j)}$ can be computed imposing the following balance

\begin{align*}
 \mathsf{E} \left[ R_i^{(j+1)} \right] &=  \mathsf{E} \left[R_i^{(j)} \right]+ N_{i+1}^{(j+1)}  - N_{i}^{(j+1)}.
 \end{align*}
 Where  $N_{i+1}^{(j+1)}$ and $N_{i}^{(j+1)}$ are respectively the expected number of symbols entering and leaving the $i$-th ripple. We have finally that
 \begin{align*}
 m^{(j+1)} p_i^{(j+1)} &=  m^{(j)} p_i^{(j)} + N_{i+1}^{(j+1)}  - N_{i}^{(j+1)}\\
 p_i^{(j+1)} &= \frac{ m^{(j)} p_i^{(j)} + N_{i+1}^{(j+1)}  - N_{i}^{(j+1)} }{ m^{(j+1)}  }.
 \end{align*}

The expected number of inactivations within step $j$ will be
\begin{equation}
n_{ \mathrm{inact} }^{(j)} = \Pr (R_{1}^j = 0) =  \left(1-p_1^{(j)}\right)^{m^{(j)}},
\end{equation}
while expected number of (overall) inactive symbols at decoding step $l$, denoted by $N_{ \mathrm{inact} }^{(l)}$, will be
\begin{equation}
N_{ \mathrm{inact} }^{(l)} = \sum_{j=1}^{l} n_{ \mathrm{inact} }^{(j)}.
\end{equation}
In the following we will adopt the shorthand $N_{ \mathrm{inact} }$ to refer to $N_{ \mathrm{inact} }^{(k)}$, that is, the expected number of inactivations required to decode.

\begin{figure}[t!]
\begin{center}
\includegraphics[width=\figw ]{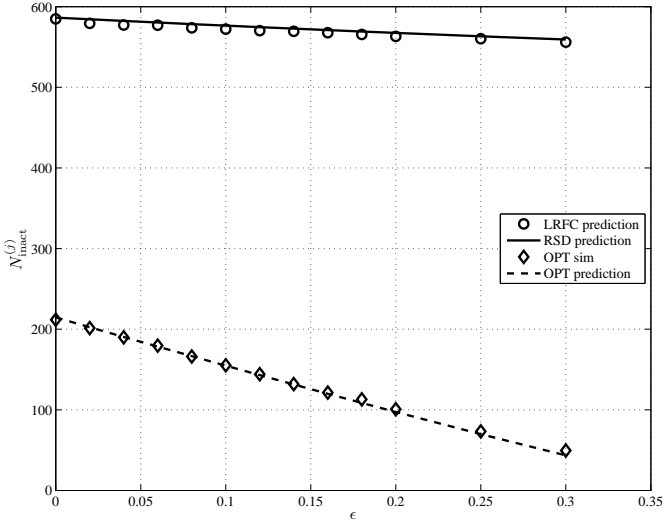}
\centering \caption{Average number of inactivations needed to decode a \ac{LRFC} and a \ac{RSD} for $k=1000$ and average output degree $\bar \Omega =12$. The markers represent simulation results and the lines represent the predicted number of inactivations for random inactivation using the proposed algorithm.}
\label{fig:k_1000_d_10}
\end{center}
\end{figure}

Fig.~\ref{fig:k_1000_d_10} shows the average number of inactivations needed to decode a \ac{LRFC}\footnote{The degree distribution of a \ac{LRFC} follows a binomial distribution.} and a \ac{RSD} with parameters $c=0.09266$ and $\delta= 0.001993$, both with average output degree $\bar \Omega =12$ and $k=1000$. It can be observed how for both distributions the estimated number of inactivations is very close to the average number of inactivations obtained through simulations.

\begin{figure}[t!]
\begin{center}
\includegraphics[width=\figw]{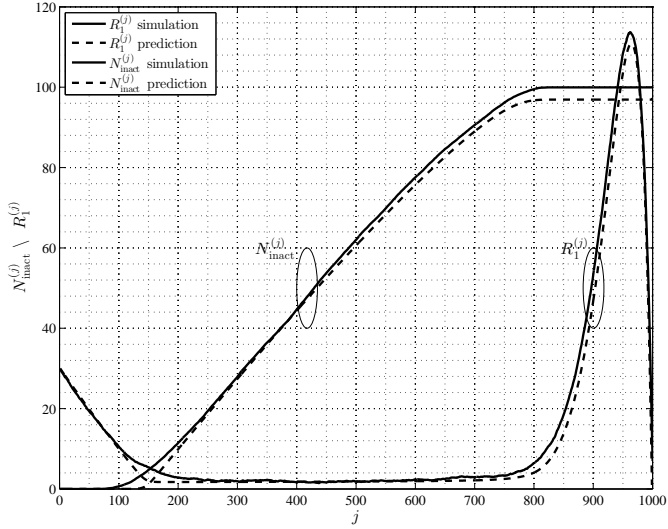}
\centering \caption{Evolution of $R_i^{(j)}$ and $N_{ \mathrm{inact} }^{(j)}$ with respect to the decoding step $j$ for a \ac{RSD} with $k=1000$ and $\epsilon = 0.2$. The solid lines represent the results of simulations and the dashed lines the prediction obtained with the proposed algorithm.}
\label{fig:ripple}
\end{center}
\end{figure}

Fig.~\ref{fig:ripple} shows the evolution of $R_i^{(j)}$ and $N_{ \mathrm{inact} }^{(j)}$ with the decoding step $j$ for the \ac{RSD} distribution at $\epsilon = 0.2$. The simulation results were obtained averaging $200$ independent realizations. It can be observed how the match between simulation results and the prediction is very tight.

\section{Degree Distribution Design}\label{sec:dist}
The algorithm proposed in section \ref{sec:model} predicts the expected number of inactivations needed to decode a \ac{LT} code. We have devised an efficient implementation of the algorithm which makes it possible to perform a numerical optimization of the output degree distribution $\Omega$.

The algorithm used to perform the numerical optimization is \ac{SA} \cite{kirkpatrick1983optimization}, a fast meta-heuristic method for global optimization. The starting point of \ac{SA} corresponds to an initial state $s_{\mathrm{init}}$ plus an initial temperature $T_{\mathrm{init}}$. At every step the temperature of the system is decreased and a number of candidate successive states for the system are generated as a slight variation of the previous state.  For high temperatures \ac{SA} allows moving the system to higher energy states but this becomes less and less likely as the temperature of the system decreases. This step is repeated until the system reaches a target energy or until a maximum number of steps are carried out. In our case the states correspond to degree distributions and the energy is a function of the predicted number of inactivations $E = f \left(N_{\textrm{inact}} \right)$.
 Note that the optimization aims at minimizing the expected number of inactivations under random inactivation which is known to be suboptimal. However, we expect that if a degree distribution $\Omega^A$ requires less inactivations than a degree distribution $\Omega^B$ under random inactivation, it will tend to require less inactivations under other inactivation strategies. Our experimental results and the experimental results in \cite{paolini2012} support this fact.
In this section we provide examples of code design based on this numerical optimization.

The goal of the optimization is minimizing the expected number of inactivations needed for decoding while complying with several design constraints. Concretely, we choose $k=10000$ and set the following constraints:
 \begin{itemize}
   \item {A target probability of decoding failure $ {P}_F^*= 10^{-2}$ at $\epsilon= 0$.}
   \item {Maximum average output degree $\bar \Omega \leq 12$.}
   \item {Maximum output degree $d_{\mathrm{max}} = 150$.}
 \end{itemize}
The first constraint is applied to the a lower bound on $P_F$ derived in \cite{schotsch:2013} and provided by
\begin{equation}
\underline {P}_F(\Omega,k,\epsilon) = \sum_{i=1}^k (-1)^{i+1} \binom{k}{i} \left( \sum_{d=1}^k \Omega_d \frac{\binom{k-i}{d}}{\binom{k}{d}}\right)^{k(1+\epsilon)}.
\label{eq:low_bound}
\end{equation}
The lower bound is tight for reception overhead slightly larger than $\epsilon=0$. This constraint aims at discarding degree distributions which may lead to excessively-high error floors.
The second and third constraints are set to control the average and maximum encoding complexity.
The metric used for optimization in this examples is $E = N_{\textrm{inact}} + f_p (\underline{P}_F)$ at $\epsilon=0$, where
\begin{equation}
f_p (\underline P_F) = \begin{cases}
0,  & \underline P_F<{P_F}^* \\
b~(1- \underline P_F/{P_F}^*,& \mathrm{else}
\end{cases}
\end{equation}
being ${P_F}^*$ the target probability of decoding failure and a $b$ a large positive number ($b= 1000$ was used in the example). The large $b$ factor ensures that degree distributions which do not comply with the target probability of decoding failure are discarded. The use of $\underline{P}_F$ in place of the actual $P_F$ stems from the need of having a fast (though, approximate) performance estimation to be used within the \ac{SA} recursion (note in fact that the evaluation of the actual $P_F$ may present a prohibitive complexity). This allows evaluating the energy of a state (i.e., degree distribution) very quickly. Although the lower bound in eq.~\eqref{eq:low_bound} may not be tight for $\epsilon=0$, where we set ${P_F}^*= 10^{-2}$, the bound indicates at which error rate the error floor of the \ac{LT} code will emerge (the bound it is very tight already for $\epsilon \approx 10^{-2}$).

\begin{figure}[t!]
\begin{center}
\includegraphics[width=\figw]{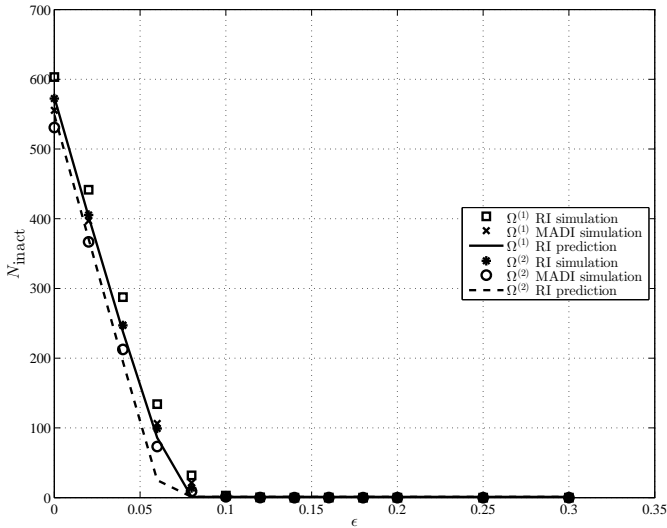}
\centering \caption{Average number of inactivations needed for decoding, $N_{\mathrm{inact}}$, for  $k=10000$. The solid and dashed lines represent the predicted number of inactivations under random inactivation for $\Omega^{(1)}$ and $\Omega^{(2)}$, respectively.  The markers denote the average number of inactivations under random inactivation and maximum active degree inactivation obtained through simulations.}
\label{fig:inact}
\end{center}
\end{figure}

\begin{figure}[t!]
\begin{center}
\includegraphics[width=\figw]{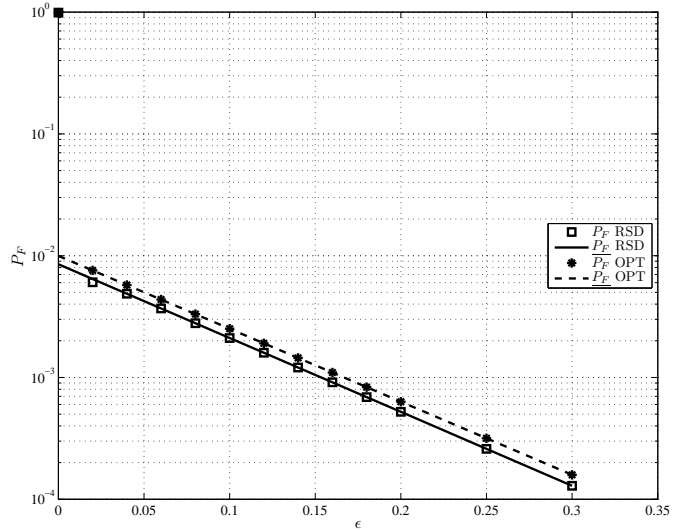}
\centering \caption{$P_F$ vs $\epsilon$ for $\Omega^{(1)}$ and $\Omega^{(2)}$. Lines represent the lower bound in Eq.~\eqref{eq:low_bound} and  markers denote simulation results.}
\label{fig:per}
\end{center}
\end{figure}

We first performed an optimization in which the degree distribution is constrained to a truncated \ac{RSD} distribution. Let $\Omega^{(R)}$  be a \ac{RSD} distribution. We define the truncated \ac{RSD} distribution, $\Omega^{(1)}$, as

\begin{equation}
\Omega_i^{(1)} = \begin{cases}
\Omega_i^{(R)},  & i < d_{\mathrm{max}} \\
\sum_{j=d_{\mathrm{max}}}^{k}\Omega_j^{(R)},  & i = d_{\mathrm{max}} \\
0,& i > d_{\mathrm{max}}.
\end{cases}
\end{equation}
Hence, the objective of this first optimization was finding the \ac{RSD} parameters $c$ and $\delta$ which minimize the number of inactivations.
In second stage we perform an optimization without any constraint on the shape of the degree distribution. We refer to the distribution obtained by this optimization method as $\Omega^{(2)}$.
Fig. \ref{fig:inact} shows  the number of inactivations needed for decoding as a function of $\epsilon$ for $\Omega^{(2)}$ and $\Omega^{(1)}$, which has parameters $c = 0.05642$ and $\delta = 0.0317$. If we look first at the results for random inactivation we can observe how the predicted number of inactivations is quite close to the actual number of inactivations obtained trough simulations. Moreover it correctly predicts the fact that  $\Omega^{(2)}$ requires less inactivations than $\Omega^{(1)}$. It is however remarkable that the truncated \ac{RSD} distribution has a very good performance in terms of number of inactivations, despite the fact that the \ac{RSD} was designed for \ac{BP} decoding and not inactivation decoding. The simulation results for maximum active degree inactivation show that, as expected, maximum active degree inactivation requires less inactivations than random inactivation, though the difference is very limited. Furthermore, $\Omega^{(2)}$ needs less inactivations than  $\Omega^{(1)}$ also under maximum active degree inactivation. For sake of completeness, the  the probability of decoding failure for $\Omega^{(1)}$  and $\Omega^{(2)}$ is provided in Fig. \ref{fig:per}.

\section{Conclusions}\label{sec:Conclusions}
We proposed a simple method to estimate the expected decoding complexity  of \ac{LT} code under inactivation decoding. The proposed method estimates the number of inactivations which have to be performed to decode an \ac{LT} code, showing to provide accurate predictions for a variety of examples.
Moreover, the model introduced in this paper has been incorporated into a numerical design procedure which allows defining output degree distributions aiming at minimizing the decoding complexity while complying with some design constraints (e.g., on the average output degree, the maximum output degree and/or probability of decoding failure). The proposed framework can be efficiently adopted to design \ac{LT} codes with various performance / complexity trade-offs under inactivation decoding.

\section*{Acknowledgement}\label{sec:ACK}
The research leading to these results has been carried out under the framework of the project `R\&D for the maritime safety and security and corresponding real time services'. The project started in January 2013 and is led by the Program Coordination Defence and Security Research within the German Aerospace Center (DLR).
\bibliographystyle{IEEEtran}
\bibliography{IEEEabrv,LT_under_ML}

\begin{thebibliography}{10}
\providecommand{\url}[1]{#1}
\csname url@samestyle\endcsname
\providecommand{\newblock}{\relax}
\providecommand{\bibinfo}[2]{#2}
\providecommand{\BIBentrySTDinterwordspacing}{\spaceskip=0pt\relax}
\providecommand{\BIBentryALTinterwordstretchfactor}{4}
\providecommand{\BIBentryALTinterwordspacing}{\spaceskip=\fontdimen2\font plus
\BIBentryALTinterwordstretchfactor\fontdimen3\font minus
  \fontdimen4\font\relax}
\providecommand{\BIBforeignlanguage}[2]{{%
\expandafter\ifx\csname l@#1\endcsname\relax
\typeout{** WARNING: IEEEtran.bst: No hyphenation pattern has been}%
\typeout{** loaded for the language `#1'. Using the pattern for}%
\typeout{** the default language instead.}%
\else
\language=\csname l@#1\endcsname
\fi
#2}}
\providecommand{\BIBdecl}{\relax}
\BIBdecl

\bibitem{byers02:fountain}
J.~Byers, M.~Luby, and M.~Mitzenmacher, ``A digital fountain approach to
  reliable distribution of bulk data,'' \emph{{IEEE} J. Select. Areas Commun.},
  vol.~20, no.~8, pp. 1528--1540, Oct. 2002.

\bibitem{luby02:LT}
M.~Luby, ``{LT} codes,'' in \emph{Proc. of the 43rd Annual IEEE Symp. on
  Foundations of Computer Science}, Vancouver, Canada, Nov. 2002, pp. 271--282.

\bibitem{shokrollahi06:raptor}
M.~Shokrollahi, ``Raptor codes,'' \emph{{IEEE} Trans. Inform. Theory}, vol.~52,
  no.~6, pp. 2551--2567, Jun. 2006.

\bibitem{shokrollahi2005systems}
M.~Shokrollahi, S.~Lassen, and R.~Karp, ``Systems and processes for decoding
  chain reaction codes through inactivation,'' Feb. 2005, {US Patent}
  6,856,263.

\bibitem{MBMS12:raptor}
{3GPP TS 26.346 V11.1.0}, ``Technical specification group services and system
  aspects; multimedia broadcast/multicast service; protocols and codecs,'' Jun.
  2012.

\bibitem{pakzad2006design}
P.~Pakzad and A.~Shokrollahi, ``{Design Principles for Raptor Codes},'' in
  \emph{Proc. 2006 IEEE Information Theory Workshop}, {Punta del Este,
  Uruguay}, Mar. 2006, pp. 165--169.

\bibitem{maneva2006lt}
E.~Maneva and A.~Shokrollahi, ``New model for rigorous analysis of
  {LT}-codes,'' in \emph{Proc. 2006 IEEE International Symp. on Inf. Theory},
  {Seattle, Washington, US}, Jul. 2006, pp. 2677--2679.

\bibitem{mahdaviani2012raptor}
K.~Mahdaviani, M.~Ardakani, and C.~Tellambura, ``{On Raptor code design for
  inactivation decoding},'' \emph{{IEEE} Commun. Lett.}, vol.~60, no.~9, pp.
  2377--2381, Sep. 2012.

\bibitem{paolini2012}
E.~Paolini, G.~Liva, B.~Matuz, and M.~Chiani, ``Maximum likelihood erasure
  decoding of ldpc codes: Pivoting algorithms and code design,'' \emph{{IEEE}
  Trans. Commun.}, vol.~60, no.~11, pp. 3209--3220, Nov. 2012.

\bibitem{kirkpatrick1983optimization}
S.~Kirkpatrick, D.~Gelatt, and M.~Vecchi, ``Optimization by simmulated
  annealing,'' \emph{Science}, vol. 220, no. 4598, pp. 671--680, 1983.

\bibitem{schotsch:2013}
B.~Schotsch, G.~Garrammone, and P.~Vary, ``{Analysis of LT Codes over Finite
  Fields under Optimal Erasure Decoding},'' \emph{{IEEE} Commun. Lett.},
  vol.~17, no.~9, pp. 1826--1829, Sep. 2013.

\end{thebibliography}
\end{document}